\documentclass[10pt]{iopart}
\usepackage{epsf}
\usepackage{fleqn}

\newcommand{\nn}{\nonumber}
\newcommand{\be}{\begin{equation}}
\newcommand{\ee}{\end{equation}}
\newcommand{\bea}{\begin{eqnarray}}
\newcommand{\eea}{\end{eqnarray}}

\newcommand{\XI}{\mbox{\boldmath $\xi$}}
\newcommand{\x}{\mbox{\boldmath $x$}}
\newcommand{\y}{\mbox{\boldmath $y$}}
\newcommand{\z}{\mbox{\boldmath $z$}}
\newcommand{\w}{\mbox{\boldmath $w$}}

\begin{document}

\title{A recurrent neural network with ever changing synapses}
\author{M Heerema \footnote[6]{E-mail address: heerema@wins.uva.nl} and WA
    van Leeuwen \footnote[7]{E-mail address: leeuwen@wins.uva.nl} }
\address{Institute for Theoretical Physics, University of Amsterdam,
  Valckenierstraat 65, 1018 XE Amsterdam, The Netherlands}

\abstract
A recurrent neural network with noisy input is studied analytically,
on the basis of a Discrete Time Master Equation.
The latter is derived from a biologically realizable learning rule
for the weights of the connections.
In a numerical study it is found that the fixed points of the
dynamics of the net are time dependent, implying that the
representation in the brain of a fixed piece of information (e.g., a
word to be recognized) is not fixed in time.

\paragraph{Keywords:} brain, recurrent neural network, biological
neural network, local learning rule, local Hebb rule, energy saving
learning rule, noisy patterns, double dynamics, adaptable synapses,
weights, Master Equation 
\endabstract

\pacs{84.35+i, 87.10+e}

\section{Introduction}
\label{introduction-synaps}

It is our purpose to construct and describe a neural net that can
learn and retrieve patterns in a way that is biologically realizable.
In an actual situation, the input to a network is noisy: the brain is
confronted with patterns that are similar, but not identical.
Therefore, we are going to model the training phase of a neural network by
considering $p$ sets, $\Omega^{\mu}$, of similar patterns $\x$,
centered around $p$ typical patterns, $\XI^{\mu}$ ($\mu=1,\ldots,p$).

In the existing literature, learning rules are used which are based on 
typical patterns $\XI^{\mu}$, and not on sets of similar patterns
$\Omega^{\mu}$.
This is biologically unrealistic: a
child does not learn `standard words', $\XI^{\mu}$, pronounced by a
`standard speaker', but hears the same word pronounced by different
speakers in different ways, i.e., the child is exposed to sets
$\Omega^{\mu}$.
In order to model biologically realistic learning, we use a learning
rule which contains patterns $\x$ belonging to learning sets $\Omega^{\mu}$
rather than the patterns $\XI^{\mu}$ alone.

We show that when this learning rule, based on noisy input
patterns is used, the network evolves to values for the strengths of
the synaptic connections, usually called `weights', that
fluctuate with respect to certain fixed asymptotic values.
For an actual brain this corresponds to the fact that the confrontation with input data leads to synaptic connections that change in
strength, through all of their lifetime, but in such a way that there
is stability in what it stores and recollects.
When a biological neural network gets as input a pattern that it has
learned a long time ago already, and which, via the tuning of the
synapses, has found a firm and fixed place on the substrate formed by
the neural tissue, it nevertheless changes the synapses.
This is not necessary of course, but can not be circumvented in an actual
biological network.
Input always changes the connections, since there is no way for an
individual neuron to know whether or not a pattern has been
encountered earlier.
All this means that the learning rule must be such that the
learning speed is not too large \cite{LNPS}: new information might otherwise
destroy too much of the old information and, hence, the network's
functioning. 
 
So far, we have been considering the situation that patterns are
presented to the network and that, through some learning rule,
synaptic connections are changed \cite{heerema}.
A next point to discuss is how an actual, biological, neural net,
while changing its connections continuously, can nevertheless
recognize a pattern. 
In other words, what is the difference, for a neural net, between
an old, i.e., already stored pattern, and a new, unknown pattern?
As noted before, also an old pattern is `stored', in the sense that it
gives rise to a change of the connections.
The answer to the question is that albeit that each individual neuron
and each synapse reacts independently of the fact whether a pattern has
been stored or not, the net as a whole reacts differently in these two
cases: see section~\ref{num}, in particular figure~\ref{figbio3}.

Neural nets have been studied with fixed \cite{DGZ,AEHW}, and with
adapting ---or dynamic--- synapses \cite{LNPS,FA,CS}.
The latter neural nets are also called nets with double dynamics
\cite{AB}.
In the context of spin glasses one speaks of coupled dynamics \cite{PCS}.
In the last part of this article, we will be concerned with neural
networks with double dynamics: see section~\ref{double}.

We study what happens in a net with ever changing
connections, by comparing what happens when a pattern is
presented to the net that has been learned before to what happens when
this is not the case.
These two cases are investigated numerically on the basis of a particular
learning rule, for which we have chosen the one we derived earlier
\cite{heerema}: it is a Mixed Hebbian-Anti-Hebbian, Hopfield like, learning
rule, which is non-symmetric with respect to post- and pre-synaptic
input, and which contains, moreover, a post-synaptic potential dependent
factor.
We found this rule assuming that building and destroying of a synapse
costs biochemical energy, and by requiring, at the same time, that the energy needed to
change a neural network be minimal.
We suppose that the patterns $\x \in \Omega^{\mu}$ that are presented
to the net (the various ways in which one and the same `word' is
presented) are chosen randomly from a set of patterns distributed
around a set of $p$ typical patterns
$\XI^{\mu}$ (the $p$ `standard words' to be learned).

Random processes can often be described in a useful way via a
so-called Master Equation for the relevant random variables \cite{kampen,LSS}.
We therefore start, in section~\ref{average-weights}, by deriving a Discrete Time Master Equation for the
random variables in question, namely the weights $w_{ij}$ of the
connections of the network.
Usually, a Master Equation is solved going from discrete time to
continuous time, which always entails some essential difficulties
\cite{HK,radons,heskes}.
Such a transition to a process that is continuous in time is often
advantageous, since a differential equation, in general, is easier to
solve than a difference equation.
In our approach, the transition to the differential equation could be
circumvented, since we had in this case at our disposal a tool that
turned out to enable us to directly solve
the difference equation itself: the Gauss-Seidel iterative method.

A question one might raise is whether a system with ever changing
connections will ever achieve some kind of stationary state.
A numerical study can not easily answer this question, since the
fluctuations of the weights are quite wild (see figure~\ref{fluc-w}). 
We therefore performed an analytic study, based on the particular learning
rule used throughout our work.
We found that the system's weights will fluctuate around certain
asymptotic values, and that the last stored pattern that has given
rise to a fixed point is roaming over $\Omega^{\mu}$, the
collection of patterns around a typical pattern $\XI^{\mu}$.
All this can be rephrased by stating that both the neural net itself
and its particular states, the fixed points, wiggle around
average values: the ever changing mind is, in some sense, stable.

The above can be summarized as follows.
In section~\ref{average-weights} we derive the Discrete Time Master
Equation for the weights of a neural network, from a learning rule,
and solve this equation analytically.
In section~\ref{num} we have two objectives: firstly, to check
numerically the analytical result of section~\ref{average-weights},
and, secondly, to study the implications of double dynamics.

\section{The weights of a network trained with noisy patterns}
\label{average-weights}

In a preceding article we derived what we have called an `energy
saving learning rule'.
When at time $t_n$ ($n=0,1,\ldots$) the weights are given by
$w_{ij}(t_n)$, and if, thereupon, a pattern $\XI=(\xi_1,\ldots,\xi_N)$
is presented to a net of $N$ neurons, then the weights are changed according to the
rule  
\be
\label{rule}
w_{ij}(t_{n+1})=w_{ij}(t_n)+\Delta w_{ij}(t_n)  \qquad  (j \in V_i) 
\ee 
where
\be
\label{global-energy-rule}
\Delta w_{ij}(t_n)= \eta_i [\kappa-
\gamma_i(\XI,\w_i(t_n))] (2\xi_i-1) \xi_j  \qquad (j \in V_i)
\ee
(see \cite{heerema}, equations ($41$) and ($42$)).
The index $n$ in $t_n$ labels subsequent moments of the net: $t_0$ is
the initial time, where the weights have there initial values, $w_{ij}(t_0)$.
In these equations $V_i$ is the collection of indices $j$ with which
neuron $i$ is connected via adaptable, non-zero synapses.
Furthermore, for the so-called stability coefficients we used the abbreviation
\be
\label{gamma-def}
\gamma_i(\x,\w_i(t)):=(\sum_{l=1}^N w_{il}(t) x_l -\theta_i)(2x_i-1) \qquad (i=1,\ldots,N)  
\ee
where $\w_i:=(w_{i1},\ldots,w_{iN})$.
The quantity arises naturally in case the dynamics of the network is
taken to be given by equation (\ref{bio-dynamics}) below; see, e.g.,
\cite{heerema}.
In the learning rule (\ref{gamma-def}) occur two quantities, $\eta_i$
and $\kappa$.
For a non-biological system they can be expressed in terms of
properties of the neural net and as a function of the patterns
$\XI^{\mu}$ to be stored in the net: for $\eta_i$, the so-called
learning rate, see equations ($42$)and ($65$)
of \cite{heerema}; for $\kappa$, the so-called margin parameter, see \cite{heerema2}. 
For biological systems, the coefficients $\eta_i$ and $\kappa$ are
replaced by suitable constants: for $\eta_i$ see \cite{heerema},
section~$6$; for $\kappa$ see \cite{heerema2}, section~$3$.

It was shown that a repeated application ($n \rightarrow \infty$)
of the rule (\ref{global-energy-rule}) gave rise to the following
expression for the weights at some ---finite or infinite--- time $t_{\infty}$
\be
\label{w-infty}
\fl
w_{ij}(t_{\infty})= \cases{w_{ij}(t_0)+N^{-1} \sum_{\mu,\nu=1}^p
[\kappa-\gamma_i^{\mu}(t_0)]  (2\xi_i^{\mu}-1) (C_i^{-1})^{\mu
  \nu}\xi_j^{\nu} & ($j \in V_i$) \\
w_{ij}(t_0)  & ($j \in V_i^c$)}
\ee
where $\gamma_i^{\mu}(t_0):=\gamma_i(\XI^{\mu},\w_i(t_0))$ and where $C_i^{\mu \nu}$ is the reduced correlation matrix, defined by
\be
\label{c-def}
C_i^{\mu \nu}:=N^{-1} \sum_{k \in V_i} \xi_k^{\mu} \xi_k^{\nu}
\ee
(see \cite{heerema} equation ($52$)).
This type of learning, and the ensuing expression for the weights,
$w_{ij}(t_{\infty})$, correspond to the idealized situation of ideal,
i.e., unperturbed, input.

\subsection{The energy saving learning rule for learning with noise}

In a realistic situation, however, the repeated training of a network will not
take place with patterns $\XI^{\mu}$ which remain exactly the same
throughout all of the training process.
Each time a certain pattern $\XI^{\mu}$ is presented to the network,
it may be slightly different.
Therefore, rather than studying what happens when patterns $\XI^{\mu}$
are presented, we will study what happens when patterns $\x$ are
learned, which belong to sets of patterns $\Omega^{\mu}$ clustered
around typical patterns $\XI^{\mu}$ ($\mu=1,\ldots,p$).
In other words, we allow the patterns to be alike ---but not
necessarily exactly equal--- to one of the $p$ representative patterns
$\XI^{\mu}$: we allow for what is called, technically, `noise'.

In a preceding article, we studied patterns $\x$, belonging to sets of
patterns $\Omega^{\mu}$ clustered around typical patterns $\XI^{\mu}$,
in the context of basins of attraction \cite{heerema2}.
In other words, we introduced $\Omega^{\mu}$ as a means to construe,
by hand, basins of attraction.
In the present article, we use $\Omega^{\mu}$ to represent noisy
patterns.
Thus the sets $\Omega^{\mu}$ appearing in the two articles have the
same meaning, namely sets of patterns around a typical pattern, but
the reasons for introducing them is different: in the preceding article
there was a mathematical motivation, whereas in the current article it
is motivated by the biological reality that patterns are never exactly
equal.

The purpose of this article is to determine the values for the
weights $w_{ij}$, in case of learning with noisy patterns.  
We start by simply conjecturing that for noisy patterns the old rule (\ref{global-energy-rule}) can essentially be
maintained: all what we do is replacing in (\ref{global-energy-rule}) the $\xi^{\mu}$'s  by $\x$'s.
Hence, we take as learning rule 
\be
\label{global-bio-energy-rule}
\Delta w_{ij}(\x,t_n)= \eta_i [\kappa-
\gamma_i(\x,\w_i(t_n))] (2x_i-1) x_j  \qquad (j \in V_i) \, .
\ee
The learning rate $\eta_i$ figuring in this expression will be
discussed in section~\ref{bio-eta}; the margin parameter $\kappa$ has been
discussed in a preceding article \cite{heerema2}. 
We will prove that this learning rule leads, on the average, to
suitable values for the weights. 
From this we conclude that the energy saving learning rule is suitable
also for learning the right patterns $\XI^{\mu}$, on the basis of wrong
(i.e. perturbed) input patterns $\x$ of $\Omega^{\mu}$.

\subsection{The Discrete Time Master Equation}
\label{markov}

When learning takes place in a biological neural network with a
learning rule of the type (\ref{global-bio-energy-rule}), we assume
that the changes
of the weights at time $t_{n+1}$ depend only on the values of the
weights $\w_i$ at time $t_n$ and the variable $\x=(x_1,\ldots,x_n)$ which is
randomly drawn out of the collection $\Omega=\cup_{\mu} \Omega^{\mu}$,
the union of disjunct sets $ \Omega^{\mu}$ of patterns $\x$
centered around typical patterns $\XI^{\mu}$, at time $t_n$.
Consequently, learning with a learning rule like
(\ref{global-bio-energy-rule}) is a Markovian process, and the weights
$w_{ij}$ are stochastic variables.
Thus we have for the new weights $w_{ij}'$:
\be
\label{recurs}
w_{ij}'= \cases{ w_{ij}+\Delta w_{ij}(\x,\w_i) & ($j \in V_i$) \\ 
w_{ij}  & ($j \in V_i^c$)}
\ee
where the $\Delta w_{ij}(\x,\w_i)$ are the increments given by the learning rule
(\ref{global-bio-energy-rule}),
\be
\label{global-bio-energy-rule-stoch}
\Delta w_{ij}(\x,\w_i)= \eta_i [\kappa-
\gamma_i(\x,\w_i)] (2x_i-1) x_j  \qquad (j \in V_i) \, .
\ee
 
Let $T_{ij}(w_{ij}'|w_{ij})$ be the probability density that a
transition takes place from the value $w_{ij}$ to the value $w_{ij}'$. 
Then we have
\be
\label{def-t}
T_{ij}(w_{ij}'|w_{ij})= \sum_{ \x \in \Omega  } p(\x) \delta(w_{ij}'-w_{ij}-\Delta w_{ij}(\x,\w_i))
\ee
where $p(\x)$ is the probability to draw $\x$ from $\Omega$.
The $\delta$-function guarantees that only transitions take place
which obey the learning rule (\ref{global-bio-energy-rule-stoch}).
Using a probability $p(\x)$ normalized to unity, i.e.,
\be
\label{norm-px}
\sum_{ \x \in \Omega  } p(\x)=1 
\ee
we find from (\ref{def-t})
the following total transition probability to a state $w_{ij}$: 
\be
\label{norm-t}
 \int \! dw_{ij}' \, T_{ij}(w_{ij}'|w_{ij}) =1 \, .
\ee
Let $P_{ij}(w_{ij},t_{n+1})$ be the probability of occurrence of the variable
$w_{ij}$ at $t_{n+1}$.
Then, the probability $P_{ij}$ and the transition probability $T_{ij}$ are related
according to
\be
\label{def-prob-p}
P_{ij}(w_{ij},t_{n+1})= \int \! dw_{ij}' \, T_{ij}(w_{ij}|w_{ij}')
P_{ij}(w_{ij}',t_n) \qquad (i=1,\ldots,N; j \in V_i) \, .
\ee
Let, moreover, the probability $P_{ij}$ be normalized according to
\be
\label{norm-p}
 \int \! dw_{ij} \, P_{ij}(w_{ij},t_n) =1 \, .
\ee
From (\ref{norm-t}) and (\ref{def-prob-p}) it follows that
\be
\label{master}
\eqalign{P_{ij}(w_{ij},t_{n+1}) - & P_{ij}(w_{ij},t_n) = \int \! dw_{ij}' \,
[ T_{ij}(w_{ij}|w_{ij}') P_{ij}(w_{ij}',t_n) \\
& -T_{ij}(w_{ij}'|w_{ij}) P_{ij}(w_{ij},t_n)]  \qquad
(i=1,\ldots,N; j \in V_i) } 
\ee
which is the Discrete Time Master Equation for the weights $w_{ij}$.

Next, let us consider the average of the weights at $t_n$:
\be
\label{av-w}
\langle w_{ij} \rangle_{t_n} := \int \! dw_{ij} \,
P_{ij}(w_{ij},t_n) w_{ij}  \qquad  (j \in V_i) 
\ee
or, using the normalization (\ref{norm-p}),
\be
\label{av2-w}
\langle w_{ij} \rangle_{t_n} = \prod_{k=1}^N \int \! dw_{ik} \,
P_{ik}(w_{ik},t_n) w_{ij}  \qquad  (j \in V_i) \, .
\ee
Using the Master Equation (\ref{master}), we obtain for the change of
the weights
\be
\eqalign{
\langle w_{ij} \rangle_{t_{n+1}} - \langle w_{ij} \rangle_{t_n} =
 & \prod_{k=1}^N \int \! dw_{ik} dw_{ik}' \,
[ T_{ik}(w_{ik}|w_{ik}') P_{ik}(w_{ik}',t_n) \\
& - T_{ik}(w_{ik}'|w_{ik}) P_{ik}(w_{ik},t_n)] w_{ij} \, .} 
\ee
Interchanging the primed and unprimed variables in the first term we
find
\be
\langle w_{ij} \rangle_{t_{n+1}} - \langle w_{ij} \rangle_{t_n}  =
\prod_{k=1}^N \int \! dw_{ik} dw_{ik}' \,
 (w_{ij}'-w_{ij}) T_{ik}(w_{ik}'|w_{ik}) P_{ik}(w_{ik},t_n) 
\ee
or, with (\ref{def-t}),
\be
\fl
\label{w-w}
\langle w_{ij} \rangle_{t_{n+1}} - \langle w_{ij} \rangle_{t_n} =
\sum_{ \x \in \Omega }
p(\x) \prod_{k=1}^N \int \! dw_{ik} \, \Delta w_{ij}(\x,\w_i)
P_{ik}(w_{ik},t_n) \, \qquad (j \in V_i) \, .
\ee
Inserting (\ref{global-bio-energy-rule-stoch}), and using (\ref{norm-p}) and
(\ref{av-w}) we find a first order difference equation for the variable
$\langle w_{ij} \rangle_{t_n}$
\be
\fl
\label{w-w-3}
 \langle w_{ij} \rangle_{t_{n+1}} -
 \langle w_{ij} \rangle_{t_n} =  \sum_{ \x  \in \Omega } p(\x) \eta_i
 [\kappa- \gamma_i(\x,\langle \w_i \rangle_{t_n})] (2x_i-1)x_j  \, \qquad
 (j \in V_i) 
\ee
which can be solved once more is known about the probability $p(\x)$.

Let $p^{\mu}(\x)$ be the probability to draw $\x$ from the disjunct collections
$\Omega^{\mu}$, normalized according to
\be
\label{norm-f}
\sum_{ \x \in \Omega^{\mu} } p^{\mu}(\x)=1 \, .
\ee
Then the probability to draw $\x$ from $\Omega= \cup_{\mu}
\Omega^{\mu}$ is
\be
\label{tot-prob}
p(\x)=\frac{1}{p} \sum_{\mu=1}^p p^{\mu}(\x) 
\ee
in agreement with the normalization (\ref{norm-px}), as may be
verified with (\ref{norm-f}) and (\ref{tot-prob}).

Let $p_i^{\mu}(x_i)$ be the probability that in the collection
$\Omega^{\mu}$ the $i$-th component of $\x$ has the value $x_i$.
Then the probability $p_i^{\mu}(x_i)$ to draw from the collection
$\Omega^{\mu}$ the vector $\x$ is given by 
\be
\label{prob-bio}
p^{\mu}(\x)=\prod_{i=1}^N p_i^{\mu}(x_i) \, .
\ee
By choosing the normalization according to
\be
\label{norm-g}
\sum_{x_i=0,1} p_i^{\mu}(x_i)=1
\ee
we find that (\ref{norm-f}) is satisfied.

We now introduce the average with respect to the set $\Omega^{\mu}$:
\be
\label{av-x}
\bar{x}_i^{\mu}:= \sum_{ \x \in \Omega^{\mu}  } p^{\mu}(\x) x_i \, .
\ee
In view of (\ref{tot-prob}), (\ref{prob-bio}), (\ref{norm-g}) and (\ref{av-x}) we have
\bea
\label{distr-x}
\sum_{ \x \in \Omega  } p(\x) x_i &=& \frac{1}{p} \sum_{\mu=1}^p
\bar{x}_i^{\mu} \\ 
 \label{distr-xx}
\sum_{ \x \in \Omega  } p(\x) x_i x_j &=& \frac{1}{p} \sum_{\mu=1}^p \bar{x}_i^{\mu} \bar{x}_j^{\mu} \, . 
\eea
With these relations we may rewrite the difference equation for the
average weights, equation (\ref{w-w-3}), in the form
\be
\label{w-w-4}
 \langle w_{ij} \rangle_{t_{n+1}} -
 \langle w_{ij} \rangle_{t_n} =  \frac{1}{p} \sum_{\mu=1}^p \eta_i
 [\kappa- \gamma_i(\bar{\x}^{\mu},\langle \w_i
 \rangle_{t_n})] (2 \bar{x}_i^{\mu} -1) \bar{x}_j^{\mu} \qquad  (j \in V_i) \, \, \,  
\ee
This equation will be solved, in the next section, in the limit of large $n$.
Note that the $\gamma_i$'s now contain the averages $\bar{\x}^{\mu}$.

\subsection{The Gauss-Seidel solution}

The equation (\ref{w-w-4}) for the average value of the weights
 $w_{ij}$ at time $t_n$, can be solved, for $n \rightarrow \infty$, in
 a way that closely parallels the method of Diederich and Opper
 \cite{opper}.
First, using (\ref{w-w-4}) recursively, we arrive at
\be
\label{opper-start}
 \langle w_{ij} \rangle_{t_{n+1}} =
 \langle w_{ij} \rangle_{t_0} + N^{-1} \sum_{\mu=1}^p
 F_i^{\mu}(t_n) \bar{x}_j^{\mu}  \, \qquad  (i=1,\ldots,N; j \in V_i)  
\ee
where
\be
\label{def-f}
F_i^{\mu}(t_n)= \frac{\eta_i}{\alpha} \sum_{m=0}^{n}
 [\kappa- \gamma_i(\bar{\x}^{\mu},\langle \w_i
 \rangle_{t_m})] (2 \bar{x}_i^{\mu} -1) 
\ee
with $\alpha=p/N$.
From (\ref{def-f}) it follows that
\be
\label{df-1}
F_i^{\mu}(t_n)-F_i^{\mu}(t_{n-1}) = \frac{\eta_i}{\alpha} [\kappa- \gamma_i(\bar{\x}^{\mu},\langle \w_i
 \rangle_{t_n})] (2 \bar{x}_i^{\mu} -1)
\ee
or, using (\ref{gamma-def}),
\be
\label{df-2}
\fl
F_i^{\mu}(t_n)-F_i^{\mu}(t_{n-1}) = \frac{\eta_i}{\alpha} [\kappa (2
 \bar{x}_i^{\mu} -1) - (\sum_{k \in V_i} \langle
 w_{ik} \rangle_{t_n} \bar{x}_k^{\mu} +\sum_{k \in V_i^c} \langle
 w_{ik} \rangle_{t_0} \bar{x}_k^{\mu} -\theta_i)] \, .   
\ee
Eliminating $ \langle w_{ik} \rangle_{t_n} $ from  (\ref{df-2}) via
 (\ref{opper-start}) yields 
\be
\label{df-3}
\eqalign{ \frac{\alpha}{\eta_i} (F_i^{\mu}(t_n)-F_i^{\mu}(t_{n-1})) =
 & [\kappa (2 \bar{x}_i^{\mu} -1) - (\sum_{k=1}^N \langle
 w_{ik} \rangle_{t_0} \bar{x}_k^{\mu} -\theta_i)]
 \\
& - N^{-1} \sum_{k \in
 V_i} \sum_{\nu=1}^p F_i^{\nu}(t_{n-1}) \bar{x}_k^{\nu}
 \bar{x}_k^{\mu} } \, .
\ee
To solve this set of linear equations, we shall rewrite them in matrix
 notation.
First of all, let us introduce a $p\times p$ matrix, $\bar{C}_i$, with matrix-elements given by
\be
\label{comp-c}
\bar{C}_i^{\mu \nu}:= N^{-1} \sum_{k \in V_i} \bar{x}_k^{\mu}
\bar{x}_k^{\nu} \, .
\ee
We will refer to this matrix as the `correlation matrix of averages'.
The connection with a usual correlation matrix becomes more apparent
 in case $\bar{x}_k^{\mu}$ can be replaced by
 $\xi_k^{\mu}$.
Then the `correlation matrix of averages' is identical to the `reduced
 correlation matrix' \cite{heerema}.
We also introduce a $p \times p$ diagonal matrix $H_i$ with diagonal
 elements given by $H_i^{\mu \mu}=\alpha/ \eta_i$.
Finally we shall denote a $p \times p$ unit matrix as $I$.
Apart from the above mentioned matrices, we introduce the vectors
 ${\bf F}_i(t_n):=(F_i^1(t_n),\ldots,F_i^p(t_n))$ and
 ${\bf G}_i:=(G_i^1,\ldots,G_i^p)$ with components $G_i^{\mu}= [\kappa (2 \bar{x}_i^{\mu} -1) -
 (\sum_{k=1}^N \langle w_{ik} \rangle_{t_0} \bar{x}_k^{\mu} -\theta_i)] $.
With these notations and abbreviations (\ref{df-3}) can be recast in
  the simple form
\be
\label{f-matrix}
 H_i \cdot {\bf F}_i(t_n) = (H_i-\bar{C}_i) \cdot {\bf F}_i(t_{n-1})+ {\bf G}_i
\ee
Solving this equation iteratively for $ {\bf F}_i(t_n)$ we obtain
\be
\label{f-it}
\eqalign{{\bf F}_i(t_n) = & [H_i^{-1} \cdot (H_i-\bar{C}_i)]^{n} {\bf F}_i(t_0)+ H_i^{-1}[I+H_i^{-1} \cdot (H_i-\bar{C}_i) \\ 
& + \ldots + [H_i^{-1} \cdot (H_i-\bar{C}_i)]^{n-1}] \cdot {\bf G}_i }
\ee
The matrix $\bar{C}_i$, as defined in (\ref{comp-c}), is
 easily seen to be positive definite and symmetric.
It then can be shown that the matrix $H_i^{-1} \cdot (H_i-\bar{C}_i)$ has  eigenvalues smaller than one \cite{seidel}.
As a consequence, we have
\be
\lim_{n \rightarrow \infty} [H_i^{-1} \cdot (H_i-\bar{C}_i)]^{n}=0 \, .
\ee
This implies that, in the limit $n \rightarrow \infty$, (\ref{f-it})
 converges to
\bea
\label{f-final}
{\bf F}_i(t_{\infty}) &=& H_i^{-1}[I-H_i^{-1} \cdot (H_i-\bar{C}_i)]^{-1} \cdot {\bf G}_i \nn\\
&=& \bar{C}_i^{-1} \cdot {\bf G}_i \, .
\eea
Substitution of (\ref{f-final}), in index-notation, into
 (\ref{opper-start}) yields the following result for the average
 weights $\langle w_{ij} \rangle_{t_{\infty}}$ $(j \in V_i)$ after a
 learning process with noisy patterns 
\be
\label{w-opper-av}
\langle w_{ij} \rangle_{t_{\infty}}=\cases{ \langle w_{ij} \rangle_{t_0} + v_{ij} & ($j \in V_i$) \\ 
\langle w_{ij} \rangle_{t_0}  & ($j \in V_i^c$)}
\ee
where 
\be
\label{v-av}
v_{ij}= N^{-1} \sum_{\mu,\nu=1}^p [\kappa
-\gamma_i(\bar{x}^{\mu},\langle \w_i \rangle_{t_0})] (2
\bar{x}_i^{\mu}-1) (\bar{C}_i^{-1})^{\mu \nu} \bar{x}_j^{\nu} \, . 
\ee
Hence, although the weights $w_{ij}(t_n)$ themselves do not converge,
since the increments $\Delta w_{ij}(t_n)$, given by
eq. (\ref{global-bio-energy-rule}), do not tend to zero for $n$ going
to infinity, the average $\langle w_{ij}(t_n) \rangle$ does.
We thus find that the actual, biological weights $w_{ij}(t_n)$ of the
neural net fluctuate around the average value $\langle w_{ij}\rangle_{t_{\infty}}$ (see figure~\ref{fluc-w}).
The expression (\ref{w-opper-av})--(\ref{v-av}) for the average weights of a network
trained with noisy patterns constitutes the main analytical result of
this article. 

In the following section we carry out a numerical analysis on the
process of learning patterns that are perturbed.
We will use a particular expression for the probability distribution
$p^{\mu}(\x)$, namely, eq.~(\ref{prob-i}), which is the same as the
one used in a preceding article \cite{heerema2}, in order to
compare the biological result (\ref{w-opper-av})--(\ref{v-av}) for the
weights and the result for the weights obtained in case of a
mathematical approach aiming at creating fixed points with prescribed
basins of attraction.

\section{Numerical analysis}
\label{num}

On the basis of the result found above, in particular equations
(\ref{w-opper-av})--(\ref{v-av}), we expect that the energy saving learning rule
(\ref{global-bio-energy-rule}) applied to a set of noisy input
patterns $\x \in \Omega^{\mu}$ will lead to satisfactory results.
By satisfactory we mean that the system can recognize patterns
belonging to the clusters of patterns $\Omega^{\mu}$.
Still more in detail we mean that, after a certain number of learning
steps, each cluster $\Omega^{\mu}$ has a fixed point, $\y^{\mu}$ say,
whereas the other patterns of $\Omega^{\mu}$ belong to the basin of
attraction of this fixed point $\y^{\mu}$.
It should be noted that when a new pattern $\z^{\mu} \in \Omega^{\mu}$
is learned, the old fixed point $\y^{\mu} \in \Omega$ can be replaced by
a new fixed point $\z^{\mu}$.
This is a direct consequence of the learning rule
(\ref{global-bio-energy-rule}) with a pre-factor given by (\ref{fac-global}), which is such that
the last learned pattern becomes automatically, and in one learning
step only, a fixed point of the dynamics, see the preceding article
\cite{heerema}.  
In this preceding article, we considered learning of $p$ patterns
$\XI^{\mu}$ ($\mu=1,\ldots,p$).
In the language of the present article, we can say that we studied, in
the preceding article, the
learning of `clusters' $\Omega^{\mu}$ ($\mu=1,\ldots,p$), each consisting of one single pattern,
$\y^{\mu}=\XI^{\mu}$.
Since there was only one pattern per cluster $\Omega^{\mu}$, the fixed point remained the same
during all of the learning process.
In this article, the situation is a little bit different: the fixed point of a
cluster $\Omega^{\mu}$ is `roaming' over $\Omega^{\mu}$, i.e., the fixed point
is no longer fixed during all of the learning process.

\subsection{A measure for the performance of a neural net}

A criterium for the way in which a neural net functions may be based
on the stability coefficients $\gamma_i$.
The more of them are positive, for given sets of typical patterns $\XI^{\mu}$, the
better the net fulfils its task of storing and recollecting patterns
\cite{heerema}. 
Inserting the final expression for the weights,
eqs. (\ref{w-opper-av})--(\ref{v-av}), into the definition of the stability
coefficients $\gamma_i$,
eq. (\ref{gamma-def}), it follows that at $t_{\infty}$ they are given by
\be
\label{gamma-av}
\gamma_i(\bar{\x}^{\mu},\langle \w_i
\rangle_{t_\infty})=\kappa  \qquad (i=1,\ldots,N; \mu=1,\ldots,p) \, . 
\ee
In view of (\ref{distr-xx}), the $\gamma$-function of the average
$\bar{\x}^{\mu}$ can be replaced by the average of
the $\gamma$-function of $\x$ itself
\be
\label{gamma-av-2}
\bar{\gamma}_i^{\mu}(\x,\langle \w_i \rangle_{t_\infty}) =\kappa  \qquad (i=1,\ldots,N; \mu=1,\ldots,p) \, .
\ee
Since the margin parameter $\kappa$ is positive, the latter equation implies that, on the
average, the $\gamma_i$ are positive.
We now recall that for a perfectly functioning network {\em all} $\gamma_i$ should be positive.
Therefore, by calculating the fraction of $\gamma_i$'s that are positive in
various cases, we can judge the quality of a neural network.

\subsection{A useful probability distribution}

We now address the question whether there exists, after a certain
number of learning steps, a (roaming) fixed point for each cluster.
An alternative way of putting this question is to ask whether there
exist $\z^1,\ldots,\z^p$ such that the $\gamma_i(\z^{\mu},\w_i(t))$ are
all positive at a certain time $t$.
We will investigate this question numerically.
To that end, we choose a particular form for the clusters
$\Omega^{\mu}$ by specifying the choice of the probability
distribution $p^{\mu}(\x)$, equations (\ref{tot-prob})--(\ref{prob-bio}).
We take for its $i$-th factor
\be
\label{prob-i}
p_i^{\mu}(x_i)=(1-b) \, \delta_{x_i,\xi_i^{\mu}}+b \,
\delta_{x_i,1-\xi_i^{\mu}} 
\ee
where $b$ is a parameter between $0$ and $1$, which we will refer
to as the `noise-parameter'.
If $b=0$ (no noise), only the patterns $\x=\XI^{\mu}$ have a non-zero
probability of occurrence.
For values of $b$ close to zero any vector $\x$ has a non-zero
probability of occurrence, but only vectors $\x$ close to one of the
$\XI^{\mu}$ have a probability of occurrence comparable to the
probability of occurrence of a typical pattern.
The particular choice (\ref{prob-i}) for $p^{\mu}(\x)$ enables us to construct the collection
of vectors $\x$ to be used as learning input vectors in our
numerical calculation.
Since in the derivation of the Master Equation the clusters $\Omega^{\mu}$ have been chosen disjunct, a vector
$\x$ cannot belong to more than one cluster.
According to the probability distribution (\ref{prob-i}), however, a
vector $\x$ which belongs to a certain cluster $\Omega^{\mu}$, has a
---very small, but--- non-zero probability to belong to any other
cluster.
This implies that (\ref{prob-i}) is not exact but only a ---very
good--- approximation to the actual situation, for which these
probabilities vanish exactly. 

The expression (\ref{prob-i}) can be used to
calculate the average $\bar{x}_i^{\mu}$.
Inserting (\ref{tot-prob}) with (\ref{prob-bio}) and (\ref{prob-i}) into
(\ref{av-x}) we obtain  
\be
\eqalign{ \bar{x}_i^{\mu} = \sum_{x_i=0,1} &
\left[ (1-b) \, \delta_{x_i,\xi_i^{\mu}}+b \,
  \delta_{x_i,1-\xi_i^{\mu}} \right] x_i \\
& \times \prod_{k \neq i} \sum_{x_k=0,1} \left[ (1-b) \,
\delta_{x_k,\xi_k^{\mu}}+b \, \delta_{x_k,1-\xi_k^{\mu}} \right] }
\ee
or
\be
\label{x-spec}
\bar{x}_i^{\mu}= (1-b) \xi_i^{\mu} +b (1-\xi_i^{\mu}) \, .
\ee
For $b=0$, the case of patterns without noise, $\bar{x}_i^{\mu}$ reduces to $\xi_i^{\mu}$.
Using this fact in our present main analytical result, given by
equations (\ref{w-opper-av})--(\ref{v-av}), one indeed recovers the old
  result (\ref{w-infty}) for the final values of the weights in case
  of noiseless input.  

It is also instructive to compare the result for the weights $\langle
w_{ij} \rangle_{t_{\infty}}$, equations (\ref{w-opper-av}),
(\ref{v-av}) with (\ref{x-spec}) and the result of the preceding
article \cite{heerema2}.
In the latter article, we calculated the weights of a recurrent neural
net, denoted as $w_{ij}(t)$, in case fixed points and basins of
attraction are taken explicitly into account, see equations
($1$)--($2$) of \cite{heerema2}. 
The only difference is found in the factor $2\bar{x}_i^{\mu}-1$, which,
in our result based upon basins, reads $2\xi_i^{\mu}-1$.
In fact, we find for the difference of the preceding and present results: 
\be
\label{diff}
w_{ij}(t) -\langle w_{ij} \rangle_{t_{\infty}}= 2\kappa N^{-1}
\sum_{\mu,\nu=1}^p (\xi_i^{\mu}- \bar{x}_i^{\mu})(\bar{C}_i^{-1})^{\mu
  \nu} \bar{x}_j^{\nu} 
\ee
where we used the expressions ($1$)--($2$) of \cite{heerema2} and
(\ref{w-opper-av})--(\ref{v-av}), together with the definitions of
$\gamma$ in the two cases.
Furthermore, we put $\langle w_{ij} \rangle_{t_0}=w_{ij}(t_0)$ for the
initial values of the weights.
Since
\be
\label{xi-x}
\xi_i^{\mu}- \bar{x}_i^{\mu}=b(2\xi_i^{\mu}-1)
\ee
the difference (\ref{diff}) is of the order $b$, i.e., small compared
to the weights themselves.
Consequently, the biological system considered here is found to be
able to realize the optimal values $w_{ij}(t)$ for the weights derived
earlier \cite{heerema2}, up to terms that are small compared to unity.
This is intriguing, since, a priori, there is no reason to expect that
a biological learning rule based on economy of energy to rebuild a
synapse \cite{heerema} will lead to values of the weights that are a
good approximation to the values found from the requirement that there
are fixed points with prescribed basins \cite{heerema2}.

Due to the fact that the final results of \cite{heerema2} and this
article for the weights are very similar, one may expect that natural,
biologically learning via the learning rule
(\ref{global-bio-energy-rule}) for noisy patterns $\x \in
\Omega^{\mu}$ ($\mu=1,\ldots,p$) will lead to larger basins of
attraction than in case of learning of noiseless patterns. 

\subsection{Storage of noisy patterns}

\begin{figure}[tbp]
\centerline{\hbox{\epsffile{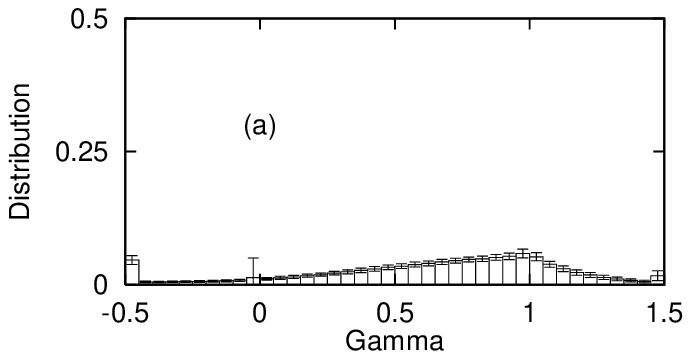} \epsffile{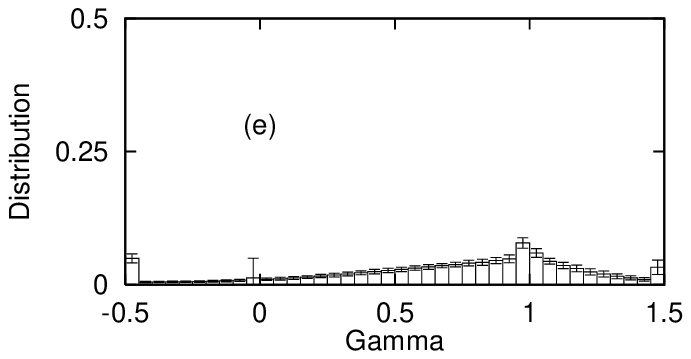}}}
\centerline{\hbox{\epsffile{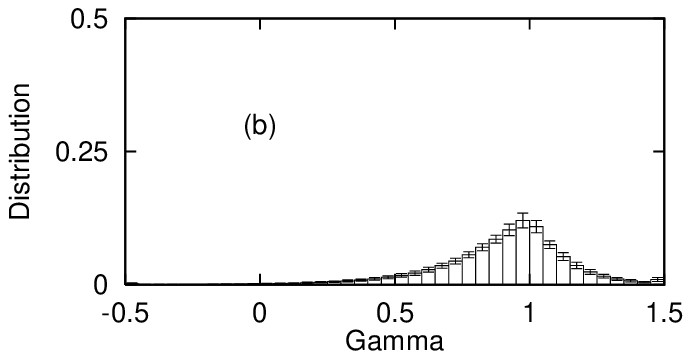} \epsffile{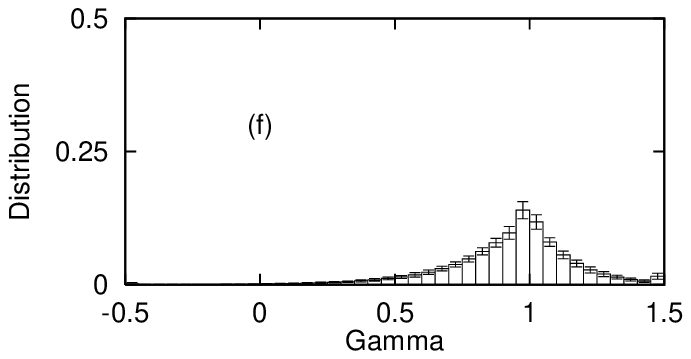}}}
\centerline{\hbox{\epsffile{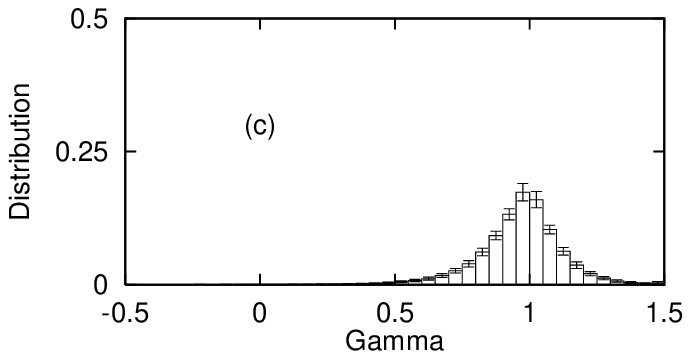} \epsffile{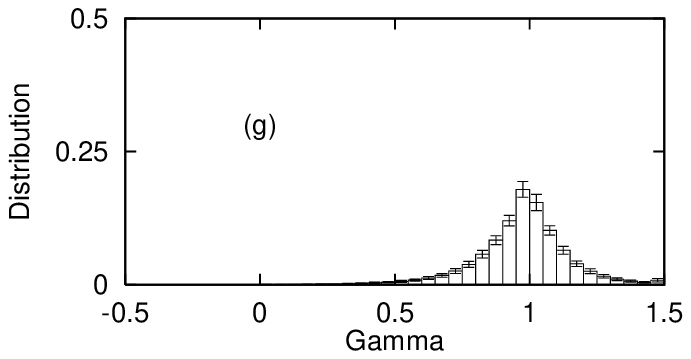}}}
\centerline{\hbox{\epsffile{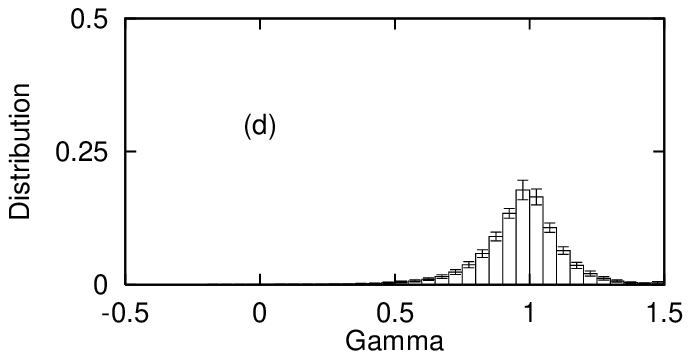} \epsffile{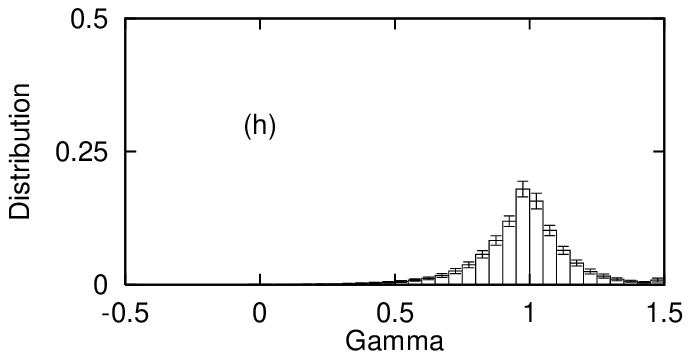}}}
\caption[0]{
\textbf{Noisy patterns}.
The performance of a neural network is measured by the stability
coefficients $\gamma_i$ related to neuron $i$, which should be positive for a
properly functioning neural net.
In the figure we have plotted averages of these $\gamma$'s, for series
of noisy input patterns, for the local learning rule (left column) and
the global learning rule (right column).
The number of learning steps increases from $32$ (top) via $160$ and
$320$ to $640$ (bottom).
The average is taken over $100$ sets of $p=32$ patterns for a neural
network with $N=128$ neurons.
The noise parameter which yields the sets $\Omega^{\mu}$ is taken to
be $b=0.01$. 
The calculations have been performed starting with a tabula rasa for
the weights: $w_{ij}(t_0)=0$, and for neurons with vanishing thresholds:
$\theta_i=0$.
The dilution in the network is $d=0.2$, the average activity of the
net is $a=0.2$.
The normalization of the weights has been fixed such that $\kappa=1$.
Both in case of global and local learning, the proposed learning rule
(\ref{global-bio-energy-rule}) leads to a very satisfactory result: almost
all $\gamma$'s become positive.   
\protect\label{figbio1}}
\end{figure}

Having chosen the sets $\Omega^{\mu}$, via $p_i^{\mu}(x_i)$,
eq.~(\ref{prob-i}), we are able to simulate a learning process with
noisy patterns.
What we will do in the numerical study below is to pick an input
vector $\x$ according to a probability distribution as given by
equation (\ref{prob-i}).
Next, we calculate the new synaptic weights $w_{ij}(t_{n+1})$,
equation (\ref{rule}), using the energy saving learning rule
(\ref{global-bio-energy-rule}).  
Finally, we calculate the $pN$ stability coefficients $\gamma_i(\z^{\mu},\w_i(t_{n+1}))$ ($i=1,\ldots,N$), where $\z^{\mu}$
stands for the last learned pattern of $\Omega^{\mu}$
($\mu=1,\ldots,p$), coefficients which we hope to be positive.
The result is shown in figure~\ref{figbio1}.

The left and right columns in figure~\ref{figbio1} correspond to two particular
choices for the factor $\eta_i$ occurring in the learning rule
(\ref{global-bio-energy-rule}), namely the `local' and `global'
learning rules, which will be described in section~\ref{bio-eta}.
Going downwards in one of these two columns, the number of learning
steps rises.
In each figure, we have put the number of gamma's as a function of its
value.
It is seen that after $300$ learning steps almost all $\gamma$'s are positive.
Hence, at the $300$-th step, most of the last learned patterns $\z^{\mu}$ are
fixed points indeed.
We note that it is instructive to compare the results of learning of
patterns with and without noise: see figure~$3$ of \cite{heerema}. 
There is almost no difference in case of the local learning rule,
whereas noise seems to diminish a little bit the effectiveness of the
(non-biological) global learning rule.

\begin{figure}[tbp]
\centerline{\hbox{\epsffile{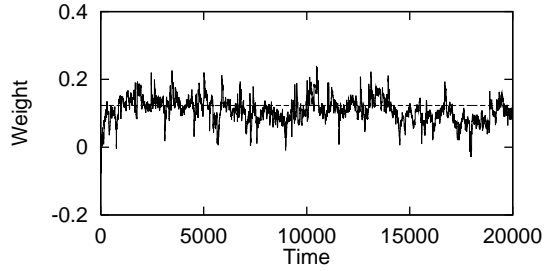}} }
\caption[1]{
\textbf{Fluctuations of the weight of a connection around its average}.
The value of an arbitrarily chosen synaptic connection $w$ is followed
during $20, 000$ learning steps, for the local learning rule.
A similar picture is obtained in case of the global learning rule.
The average value $\langle w  \rangle_{t_{\infty}}$ as predicted by equations
(\ref{w-opper-av})--(\ref{v-av}) is depicted as a horizontal dashed line.
It is observed that the value of the connection fluctuates
around the dashed line. 
The same parameters are chosen as in figure~\ref{figbio1}, i.e., $N=128$, $p=32$, $a=0.2$, $b=0.01$, $d=0.2$,
$\kappa=1$, $w_{ij}(t_0)=0$ and $\theta_i=0$ for all $i$ and $j$.
\protect\label{fluc-w}}
\end{figure}

Finally, to illustrate the fact that the weights of the synaptic connections fluctuate around the
average value as given by the expression (\ref{w-opper-av})--(\ref{v-av}), we have
plotted, in figure~\ref{fluc-w}, the time-evolution of the weight of
an arbitrarily chosen connection together with its average value.

\subsection{The learning rate $\eta_i$}
\label{bio-eta}

So far, we have been concerned with learning, and problems related to
learning.
In our study of the learning process, the weights $w_{ij}$ of the
synaptic connections changed according to the learning rule (\ref{global-bio-energy-rule}),
in which a factor $\eta_i$ occurred, the so-called learning rate,
which, so far, was left unspecified.
In this subsection, we focus the attention on this factor $\eta_i$.
In a preceding article, we showed that in a process of ideal learning, i.e., such that the energy needed to
change the synapses is minimal, the learning rate $\eta_i$ was given by
\be
\label{fac-global}
\eta_i=1/ \sum_{k \in V_i} x_k \, .
\ee
Depending on neuron activity not restricted to two neurons only, this factor is
non-local and therefore biologically unrealistic (see \cite{heerema},
section~$6$).
In a biological context, it should be replaced by some local approximation,
for instance, a constant like
\be
\label{fac-local}
\eta_i=1/ Na  \qquad (i=1,\ldots,N) 
\ee
where $a$ denotes what is called the mean-activity,
i.e., it is the probability that an arbitrary neuron $i$ is in the state $+1$. 
In figure~\ref{figbio1}, the left column of pictures corresponds to the
biological, local learning rule, i.e., equation
(\ref{global-bio-energy-rule}) with $\eta_i$ given by
(\ref{fac-local}), while the right column corresponds to the global
learning rule, i.e., equation (\ref{global-bio-energy-rule}) with
$\eta_i$ given by (\ref{fac-global}).

\subsection{Retrieval by a biological network with ever changing connections}
\label{double}

Is there more one can say on the value for the learning rate $\eta_i$ in case of a
biological network than that it should be an approximation to the
value (\ref{fac-global}), which guarantees that the process of
learning takes place in an energetically most economical way?
The answer is affirmative in case one requires that the learning is
good enough to store patterns, but is not that good that it stores
each learned pattern in only one learning step, as is the case for the
global learning rule, i.e., the learning rule with $\eta_i$ given by
equation (\ref{fac-global}).
In other words, it will turn out in this section that it is of
advantage to learn via a learning rule that is not able to always
store a new pattern in only one learning step.
The reason for this counter-intuitive requirement which we are going
to impose, is a consequence of the fact that we demand that the
network be able to retrieve patterns {\em and} change connections at the
same time.

In most models of neural networks one distinguishes between a
learning phase and a retrieval phase.
In the learning phase the weights are changed according to some rule,
in the retrieval phase the weights are kept fixed.
In a biological neural network such a separation of phases does not
occur.
Weights do not stop changing in the retrieval phase when a stimulus is
presented, and this is precisely what is happening when a neural
network has to recognize a pattern.
If the change due to the stimulus would be too close to the `ideal'
value (\ref{fac-global}), the network would change in such a way that
every new pattern would immediately be learned, and, hence, be recognized.
And this is not what should be the case: if every new pattern would be
stored immediately, it could not easily be
distinguished from a pattern that had been stored in the network a long
time ago already.
Therefore, we must require that $\eta_i$ is sufficiently unequal to
the value (\ref{fac-global}), which it has in case of the global
learning rule.
If we take $\eta_i$ larger and larger with respect to the value (\ref{fac-global}), network changes will become too
large for the network to function properly \cite{LNPS}.
So we are left with the possibility that $\eta_i$ has a value
somewhere between zero and the value (\ref{fac-global}), which is
large enough to store patterns, and small enough for the network to
distinguish between new and formerly learned patterns.

\begin{figure}[tbp]
\centerline{\hbox{\epsffile{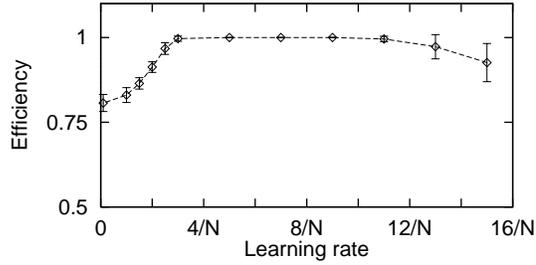}} }
\caption[2]{
\textbf{Storage efficiency for local learning}.
The neural net has been initialized by taking a tabula rasa for the
weights, followed by some learning process.
Then, the fraction of positive $\gamma$'s, after one more learning
step, has been plotted as a function of the learning rate $\eta$.
For a biological network with ever changing connections $\eta$ should
not be too effective, i.e., not in the range $3/N$ to $11/N$.
We took a neural net with $N=512$ neurons, we used $20$ initial
learning steps.
Furthermore, we took thresholds $\theta_i=0$, dilution $d=0.2$,
activity $a=0.2$ and noise level $b=0.01$.
\protect\label{figbio2}}
\end{figure}

The above qualitative statements should now be made quantitative.
In figure~\ref{figbio2} we consider the storage of one pattern.
We have plotted the percentage of positive $\gamma$'s as a function of
the learning rate $\eta$.
For $\eta$ in the range ($3/N$,$11/N$) all, or almost all, $\gamma$'s are
positive after one learning step.
For $\eta \approx 1/N$, only $80 \%$ of the $\gamma$'s are positive
after one learning step.  
We conclude from all this that the factor $\eta$ figuring in the
learning rule (\ref{global-bio-energy-rule}) should be of the order of
$3/N$ or less.
Such a value guarantees that a biological network, which is bound to
change its connections also during retrieval, does not learn so fast
that it recognizes patterns already after one learning step, as in the
case of the global learning rule (\ref{fac-global}).

\begin{figure}[tbp]
\centerline{\hbox{\epsffile{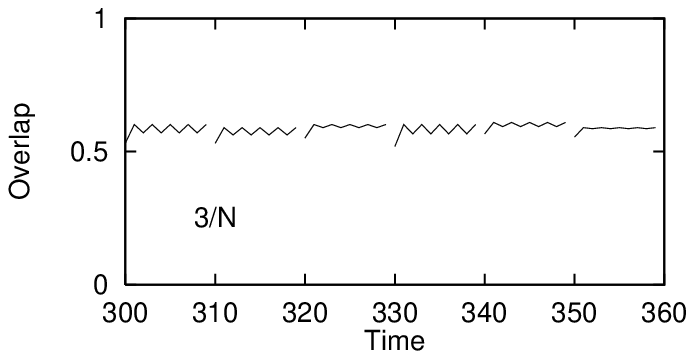} \epsffile{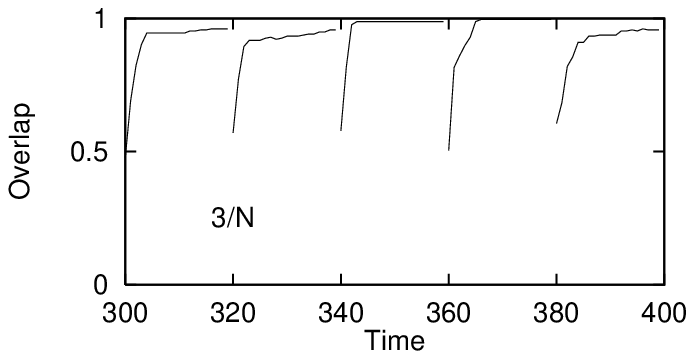} }}
\centerline{\hbox{\epsffile{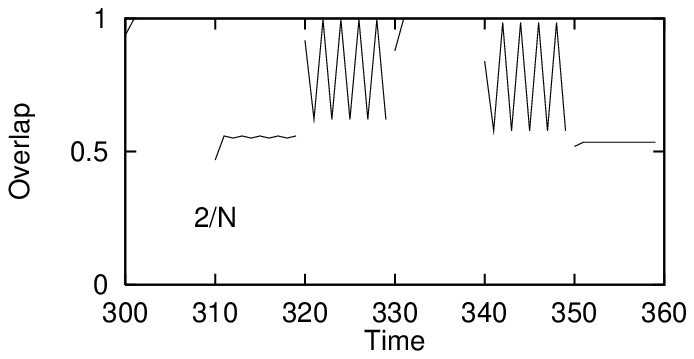} \epsffile{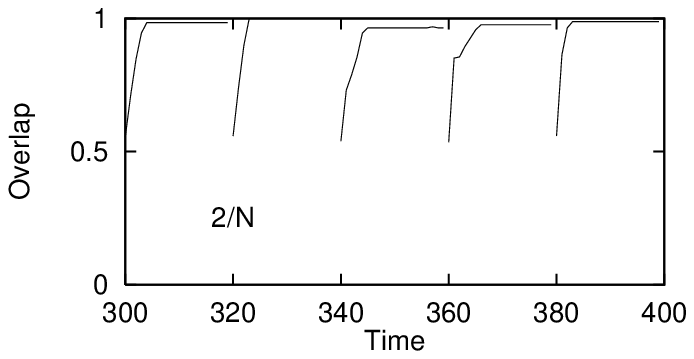} }}
\centerline{\hbox{\epsffile{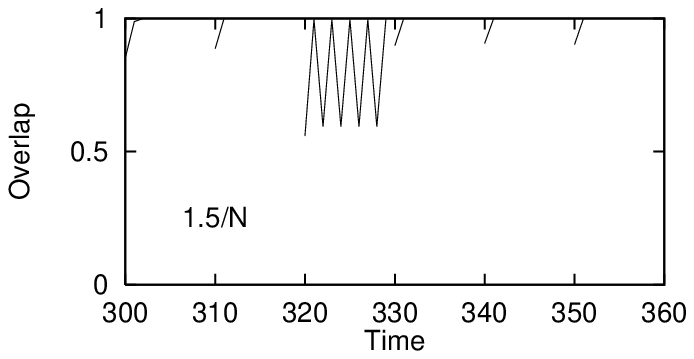} \epsffile{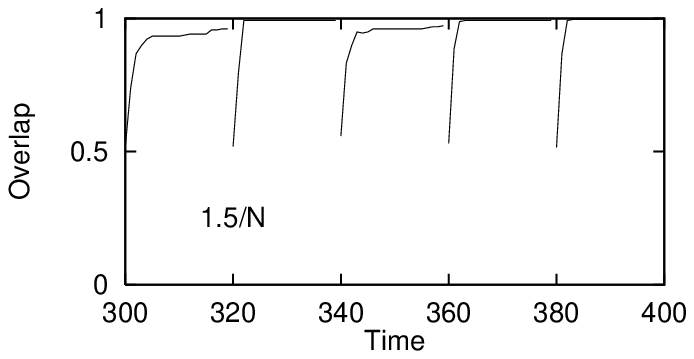} }}
\centerline{\hbox{\epsffile{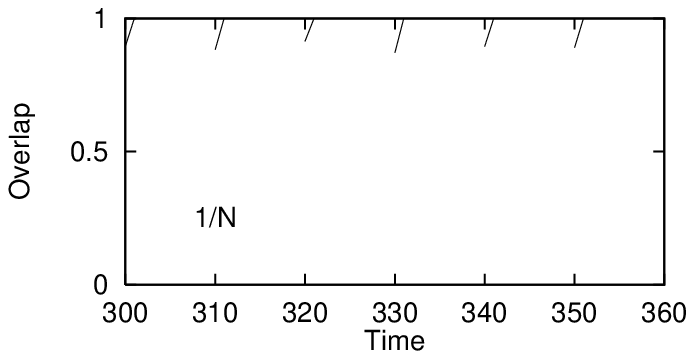} \epsffile{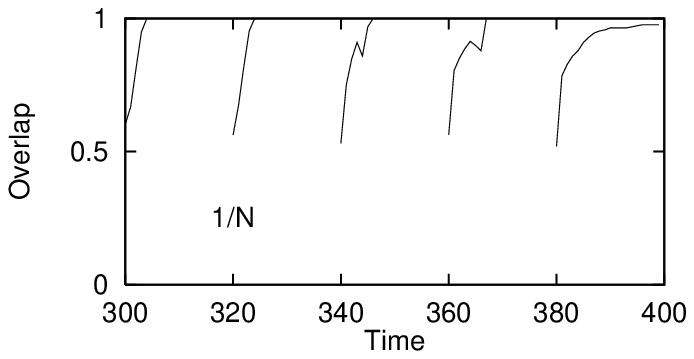} }}
\caption[3]{
\textbf{Retrieval of learned and random patterns}.
In each of the pictures at the left, a learned pattern is
  presented six times to a network, for $\eta=3/N$, $2/N$, $1.5/N$ and
  $1/N$ (top to bottom).
The recognition is preceded by a learning stage which took $300$
  learning steps. A pattern is recognized when the overlap is $1$.
In each of the pictures at the right, a random picture is presented
  five times to the network, after a learning period of the same
  number of $300$ time-steps.
Again $\eta$ varied from $3/N$ to $1/N$ (top to bottom).
A pattern is seen to evolve almost always to some other, stable,
  pattern, for all values of $\eta$, since the overlap with the
  preceding pattern almost always tends to $1$.
The network considered had the following network properties.
Number of neurons $N=512$, number of patterns $p=16$, mean activity
  $a=0.2$, dilution $d=0.2$, noise parameter $b=0.05$.
The parameter $\eta$ occurs in the learning rule (\ref{global-bio-energy-rule}) used to
  perform the updating of the weights of the connections.   
\protect\label{figbio3}}
\end{figure}

In order to compare what happens when an already learned pattern is
presented to the network with what happens in the net when a totally
new pattern is the input, it is useful to define the overlap function of
the two input patterns. 
The `overlap', $Q(\x,\y)$, of two binary patterns $\x$ and
$\y$ of $N$ bits defined in the usual way, is given by
\be
Q(\x,\y)=\frac{1}{N} \sum_{i=1}^N (2x_i-1)(2y_i-1) \, .
\ee
If $x_i=y_i$ for all ($i=1,\ldots,N$), the overlap takes its maximal
value $+1$; if $x_i=1-y_i$, for all $i$, the overlap takes its minimal
value $-1$.
In figure~\ref{figbio3} we compare the functioning of a neural network
that changes its connections during the process of `recognition' of
previously learned patterns (left column) and random, non-learned,
patterns (right column), for values of $\eta$ going down from $3/N$ (top)
to $1/N$ (bottom).
In the left column we have plotted, vertically, the overlap
$Q(\x(t_n),\z^{\mu}(t_n))$ of an arbitrary learned pattern $\x(t_n) \in
\Omega$ that is presented to the net, and the last learned fixed point
$\z^{\mu}$.
Both may change permanently in time because of the continuous updating of
the weights.
In the right column we have plotted, vertically, the overlap
$Q(\x(t_n),\x(t_{n+1}))$ of an arbitrary pattern $\x(t_n)$, not
previously presented to the network and the pattern $\x(t_{n+1})$
generated by the network one time step later.
In both columns, time steps are plotted along the horizontal axis.

In our numerical study, the connections $w_{ij}(t_n)$ are changed
according to the rule (\ref{rule}) combined with the learning rule 
(\ref{global-bio-energy-rule}); the patterns $\x(t_n)$ are updated
using the usual dynamics \cite[page $20$]{amit}
\be
\label{bio-dynamics}
x_i(t_{n+1})=\Theta_{\rm H}( \sum_{j=1}^N w_{ij}(t_n)x_j(t_n)
-\theta_i)  \qquad (i=1,\ldots,N) 
\ee
applied parallelly, i.e., at a time $t_n$, all neurons $i$ update
their states $x_i(t_n)$ simultaneously.
The learned patterns $\x$ are chosen according to the probability
$p^{\mu}(x)$ around $\XI^{\mu}$, equation (\ref{prob-i}).
The arbitrary, non-learned, patterns $\x$ are chosen randomly with mean activity $a=0.2$.

In each of the four pictures of the left column learned patterns are
presented to the network, and followed during ten learning steps.
In the top left picture no recognition takes place, whereas for lower
values of $\eta$ the recognition capability of the network rises.
In the bottom left picture recognition always takes place.
In case recognition takes place for an $\x \in \Omega^{\mu}$, it is
found that the fixed point $\z^{\mu}$ does not change in time.
If, however, recognition does not occur, the fixed point $\z^{\mu}$ was
found to change in time.
Observe that in some cases the learned patterns seem to evolve to a
two-state attractor, in contrast to what one might expect.
In fact, we showed in article \cite{heerema} that when a new pattern is
learned with the global form of learning rule
(\ref{global-energy-rule}), one arrives at a one-state attractor
(fixed point) after one learning step: see the first new paragraph of
\cite{heerema} under equation ($43$). 
Hence, we may expect that in the left column, where we use a local
learning rule that approximates the global one, one-state attractors
would occur only.
The occurence of two-state attractors can only be a consequence
of the fact that, in contrast to the treatment of \cite{heerema}, the
weights always change in time and/or the approximation of the global
learning rule by a local learning rule.

In case random patterns are the input to the network, there is in
general no evolution of the network to one of the fixed points $\z^{\mu}$: a
numerical study of the overlap of a random pattern and any of the
$\z^{\mu}$ turned out to yield an overlap which was always less
than $1$.
What we have pictured in the right column of figure~\ref{figbio3} is
evolution of a random pattern during $20$ time steps.
For all values of $\eta$ a random pattern evolves to a pattern that
remains stable or almost stable under the network dynamics.
In related cases (see, e.g., \cite{AGS} and \cite{amit} section
$4.1$) these `spurious states' are found to vanish when the dynamics
of the network is taken to be stochastic rather than deterministic.

Not only in case of learned patterns but also in case of random
patterns fixed points $\z^{\mu}$ have been found which change in time.
This is due to the fact that the weights change continuous in time or
due to the approximation of the global learning rule by a local learning rule.
We do not pursue this and other points related to figure~\ref{figbio3}
any further. 

\section{Conclusions}

The basis of this article is the learning rule for noisy patterns,
equation~(\ref{global-bio-energy-rule}).
We found, by a numerical study of this learning rule, that storage of
noisy patterns leads to fixed points that move around in collections
$\Omega^{\mu}$ that are representative for the noisy patterns.
An analytical study of the same learning rule reveals that the weights
found via this rule fluctuate, as long as learning or retrieval of
patterns takes place, around certain average values, for which the
explicit expression given by (\ref{w-opper-av})--(\ref{v-av}) could be derived.

In the limit of vanishing noise in the input, we recover the
expressions for the weights obtained earlier \cite{heerema} on the basis of a totally
different approach, namely, economy of energy in case of synaptic change.
This is satisfactory, because it yields an
independent check of their correctness.

A comparison with other results obtained earlier \cite{heerema2}, in which we
determined the optimal weights for a neural net with prescribed basins
of attraction, shows that the biological updating rule of the present
article, eq. (\ref{global-bio-energy-rule}), realizes the
latter results via eqs. (\ref{w-opper-av})--(\ref{v-av}) up to terms
of the order of the noise parameter $b$, which are small compared to
one.

\end{document}